\documentclass[12pt]{article}
\usepackage{eepic}    
\usepackage{epsfig}
\usepackage{amsmath}
\usepackage{makeidx}	     
\usepackage{graphicx}	     

\usepackage{multicol}	     
\usepackage{latexsym, graphicx, epsfig}

\newcommand{\ee}{\end{equation}}
\newcommand{\be}{\begin{equation}}
\def\ba{\begin{eqnarray}}  \def\ea{\end{eqnarray}}
\def\si{\sigma} 
\def\t{\tau} \def\r{\rho}  \def\f{\phi}
\def\x{\xi}   \def\D{\Delta} \def\th{\theta}
 \def\y{\eta} \def\p{\pi}  
\def\a{\alpha}  \def\g{\gamma} 
\def\G{\Gamma}   
  
\def\h{{\rm h}} 
\def\2{\frac{1}{2}}\def\ra{\rightarrow} \def\Ra{\Rightarrow}
   
\def\lra{\leftrightarrow}
\newcommand{\bth}{\begin{theorem}} \newcommand{\eth}{\end{theorem}}
\newcommand{\bpr}{\begin{proof}}
\newcommand{\epr}{\qed \end{proof}}
\newcommand{\ben}{\begin{enumerate}} \newcommand{\een}{\end{enumerate}}
\newcommand{\bit}{\begin{itemize}} \newcommand{\eit}{\end{itemize}}

\begin{document}
\title{The Initial Value Problem for Wave Equation and a Poisson-like Integral in
Hyperbolic Plane }
\author{Francesco Catoni and Paolo Zampetti\footnote{e-mail: paolo.zampetti@enea.it}
\vspace{4mm} \\ \it ENEA, C. R. Casaccia; \vspace{1mm} 
\it 
00123  S.Maria di Galeria, Roma, Italy}
\date{\today}
\maketitle
\begin{abstract}
In recent time, by working in a plane with the metric associated with
wave equation (the Special Relativity non-definite quadratic form),
a complete formalization of space-time trigonometry and a Cauchy-like integral
formula have been obtained.

In this paper  the concept that the solution of
a mathematical problem is  simplified by using a ``mathematics'' with
the symmetries of the problem, actuates us for studying the wave
equation (in particular  ``the initial values problem'') in a plane
where the geometry is the one ``generated'' by the wave equation
itself. 

In this way, following a classical approach, we point out the well known differences with
respect to Laplace equation notwithstanding their formal equivalence (partial
differential equations of second order with constant coefficients) and
also show that the same conditions stated for Laplace equation
allow us to find a new solution. In particular taking as  "initial data" 
for the wave equation an arbitrary function given on an arm of an equilateral hyperbola,
a "Poisson-like"  integral formula holds.  \\

\noindent{\bf Keywords:} Hyperbolic geometry. Wave equation. Boundary value problem. \\
{\bf PACS:} 02.60.Lj (Ordinary and partial differential equations; boundary value problems)
\end{abstract}

\section{Introduction}
In a recent paper \cite{noi2} and books \cite{libro}, \cite{booklet}
it has been shown how a complete formalization of Minkowski's space-time
geometry and trigonometry has been obtained by means of hyperbolic numbers that, 
from an algebraic point of view, are the simplest  extension of complex numbers and are defined as
$$\{z=x+\h\,y; \,\,\h^2=1\,;\,\,x,y \in {\bf R} ;\; h\notin {\bf R}\}.$$
These results have been obtained working in a
Cartesian plane with the non-definite metric corresponding to the modulus
of hyperbolic numbers (square distance: $d^2=x^2-y^2$), so as the Euclidean
distance corresponds to modulus of complex numbers \cite{booklet}.
Here we call {\sf hyperbolic} the plane with this metric.

By means of this approach, in \cite{libro} the bases
for studying the functions of a hyperbolic variable have been set and, for these functions,
a Cauchy-like integral formula has been stated \cite{Cl}.

Since these results refers to functions satisfying the two-dimensional wave equation here we begin
 to investigate if for studying this equation, in particular with regard to the initial value problem, it is appropriate to work in the hyperbolic plane.\\
Practically we came back from the results of Special Relativity in the following meaning:\\
{\sf Invariance of wave equations $\Ra$ Lorentz transformations $\Ra$
Minkowski (hyperbolic) geometry $\Ra$ wave equation in hyperbolic plane.}\\
																				In this approach, save for the recalled novelty, we follow Riemann who
obtained his integral formula \cite[p. 450]{ch} as a precursory
of special relativity in the meaning that for studying the initial
value problem for the wave equation, he considered as equivalent
space and time and represented them as coordinates in a plane.

In the appendix
we ''translate'' in the hyperbolic plane some properties that hold in Euclidean geometry.
\section{Initial Data Problem for Wave Equation}\label{IDPr}
Following the Cauchy theory \cite{ch}, the solution of a partial differential  equation (PDE) of degree $N$
can be obtained by a series development around a point in which the values of the
function and its partial derivative, up to degree $N-1$, are given.\\
As an exception to this approach, the solution of  the second degree
partial differential Laplace equation,	is determined by just the values of one
arbitrary function given on the frontier of a
domain with ``appropriate regularity conditions'' \cite[Chap. IV]{ch}.
This problem is known as Dirichelet's problem \cite[Chap. IV \S  2]{ch}.

As a difference from Laplace equation for which the solution is determined inside a closed domain,
 for the wave equation the initial data are given on an open, appropriate curve \cite{ch} and
the solution is determined in the two opposite sides of the given curve, in particular in the domain
 determined \cite[p. 450]{ch} by the parallel to axes bisectors (characteristic lines) from the extreme points of the curve. \\
In this paper we follow the Euclidean approach to Laplace equation and translate it
to a ``hyperbolic approach'' to wave equation. We see that, studying the
problem in this way, a new situation arises. 
In particular, by means of a ``Poisson-like'' integral,
we obtain the sum of the solutions in two points that we define as symmetric by extending
to hyperbolic plane the well known symmetry with respect to a circle in Euclidean plane
\cite[Sect. 7.5]{libro}. 

We begin by translating the basic mathematics into hyperbolic geometry.
\subsection{Integral Identity in Hyperbolic Plane }	 \label{IH}
For studying the initial data problem for partial differential equations,
the Green's formulas are usually applied \cite{ch}. They represent a particular
application of Gauss formula
\be
\int_D\int\left(\frac{\partial X}{\partial x}+\frac{\partial Y}{\partial y}
\right)\;dx\,dy=\oint_\G (X\,dy-Y\,dx)		 \label{iGreen}
\ee
that transforms the integral on a domain into an integral on its frontier.  

Let us now apply this identity to wave equation and look for its application to the ``initial value
 problem'' following the Green's approach. \\
 Let us consider two arbitrary functions $u,\,v$
continuous, with the derivatives that appear in the formulas, in a domain
$D$ up to its contour $\G$ \cite[Chap. IV]{ch}. Let us set
\be X=v\frac{\partial u}{\partial x},\,\,Y=-v\frac{\partial u}{\partial
y} , \label{iGreenA}
\ee
and introduce the differential parameters \cite[Chap 8]{libro}
in the flat hyperbolic plane, i.e., with pseudo-Euclidean metric
\be
\mbox{ ( a ) }\qquad\D_2\,u=\frac{\partial^2 u}{\partial x^2}-
\frac{\partial^2 u}{\partial y^2}\;\;,\qquad
\quad \mbox{ ( b ) }\qquad\D(v,\,u)= \frac{\partial v}{\partial x}\;\frac{\partial u}{\partial x}-
\frac{\partial v}{\partial y}\frac{\partial u}{\partial y}\;.
\label{iGreenB}
\ee
Equation (\ref{iGreen}) becomes
\be
\int_D\int v\,\D_2\, \,u\;dx\,dy+ \int_D\int \D(v,\,u)\;dx\,dy=\oint_\G
v\,\left(\frac{\partial u}{\partial x}\;d\,y+\frac{\partial u}{\partial y}\;d\,x
\right) \label{ah1}.
\ee
Now we go on as it is usually done for studying the ``initial values
problem'' for Laplace equation. \\
By subtracting from Eq. (\ref{ah1}) the equation obtained by changing
$v\lra u$, we obtain a {\it Green identity} for the differential
operators (\ref{iGreenB} a) 
\be
\int_D\int\,(v\,\D_2\,u-u\,\D_2\,v)\;dx\,dy=\oint_\G\left[
v\,\left(\frac{\partial u}{\partial x}\;d\,y+\frac{\partial u}{\partial y}\;d\,x
\right)-u\,\left(\frac{\partial v}{\partial x}\;d\,y+\frac{\partial v}{\partial y}\;d\,x
\right)\right] . \label{ah444}
\ee
In order to calculate the line integral along the curve $\G$, on the right hand side
of Eq. (\ref{ah444}), a local reference frame is used. This frame has its origin
in the point that moves along the curve and an axis ($\t$) tangent to the curve,
oriented according with the integration direction. \\
For the construction of the other local axis $n$, the hyperbolic geometry is applied
to the plane $x,\,y$. So we take the $n$ axis in the direction of the
hyperbolic normal to $\t$ axis and  oriented so that the frame $n,\,\t$ is congruent
with the orientation of $x,y$ frame. 
According with the topology of hyperbolic plane \cite{noi2},
different kinds of pairs of unity vectors  originate. \\
A detailed treatment of this subject, based on \cite{noi2}, is developed in App. \ref{ntau} and
 summarized in the caption of Fig. \ref{fig1} where the hyperbola with $|dy/dx|>1$, that is the 
 curve employed in this paper,	is considered.	  \\
Now we write 
the argument in brackets of the integrals in the right hand side of Eq. (\ref{ah444}) as a function
of the local coordinates $n,\,\t$.  \\
Therefore, by setting \bit
\item
${\partial u}/{\partial n}$ the derivative of $u$ in the direction of $n$; \eit
taking into account that
\bit
\item in the line integration it results $d\,n=0$,
\item the relations between the derivatives with respect to the orthogonal directions
of the tangent and the ``hyperbolic normal'' to a curve are (Eq.  (\ref{lorecr}))
\be \frac{\partial\,x}{\partial\,\t}= \frac{\partial \,y}{\partial \,n} \;\;,\;\;
\frac{\partial\,x}{\partial \,n}= \frac{\partial \,y}{\partial \,\t}, \ee
\eit
we have
\be
\frac{\partial u}{\partial x}\;d\,y+\frac{\partial u}{\partial y}\;dx \equiv
\left(\frac{\partial u}{\partial x}\;\frac{\partial
y}{\partial \t}+\frac{\partial u}{\partial y}\frac{\partial x}{\partial \t}
\right)\; d\t\equiv
\left(\frac{\partial u}{\partial x}\;\frac{\partial
x}{\partial n}+\frac{\partial u}{\partial y}\frac{\partial y}{\partial n}
\right)\; d\t\equiv \frac{\partial u}{\partial n}\;d\t\,.  \label{iGreenC}
\ee
With these definitions and expressions Eq. (\ref{ah444}) becomes
\be
\int_D\int\,(v\,\D_2\,u-u\,\D_2\,v)\;dx\,dy=
\oint_\G\left(v\,\frac{\partial u}{\partial n}-
u\,\frac{\partial v}{\partial n}\right)\;d\t. \label{ah4}
\ee
Therefore, by working in hyperbolic plane, i.e., with the geometry related with
wave equation \cite{libro}, {\it the same integral identity} (\ref{ah4}) {\it
that holds by applying Euclidean geometry to $x,y$ plane in the study of
Laplace equation} \cite[p. 252, Eq. (26)]{ch}, {\it has been obtained.}   \\
Now we go on as it is usually done for studying the ''initial values problem"
for Laplace equation. \\
In particular if $u,\,v$ satisfy the wave equation, the left-hand side of Eq. (\ref{ah4})
is zero and we have
\be
\oint_\G\left(v\,\frac{\partial u}{\partial n}-u\,\frac{\partial v}{\partial n}
\right)\;d\t=0 .\label{ah5}
\ee
By setting in Eq. (\ref{ah4}) $v=1$ we obtain
\be
\int_D\int \D_2\,u\;dx\,dy=\oint_\G \frac{\partial u}{\partial n}
\;d\t, \label{ah2}
\ee
and if the function $u$ satisfy in $D$ the wave equation and
holds the appropriate regularity conditions, that for Laplace equation
are: to be continuous with
the partial derivatives in the domain and satisfy the H\"{o}lder
conditions on the contour \cite{mf}, \cite[p. 50]{ST}, we have:

{\it The integral on the contour} ($\G$) {\it of the derivative of
$u$ with respect to the normal is zero}:
\be \oint_\G \frac{\partial u}{\partial n}\;d\t=0\;. \label{ah22}\ee   \\
\subsection{Characteristic Domains in Hyperbolic Plane}
\label{GLWE} \label{GREEN-La}
In the application of the relations of the previous section
to the ``initial value problem'' for Laplace equation and in complex
analysis for demonstrating the Cauchy integral formula, the singularity at a point
$Q\equiv(x,\,y)$ is excluded from the domain of integration by means
of a circle, centered in $Q$, with radius $r\ra 0$. In Euclidean plane
this  circle represents the locus of points at the same
distance from $Q$ and allows one to obtain useful simplifications and relevant results. \\
In hyperbolic plane the locus of points at the
same distance from a given point is the equilateral hyperbola, then in
\cite{Cl} the circle with radius $r\ra 0$ has been replaced, by
an equilateral hyperbola with semi-diameter $\r\ra 0$.
This means that the hyperbola becomes the parallel to axes bisectors from
the point $Q$. \\
This domain can be recognized as the one considered by Riemann for
his approach to the initial value problem for wave equation 
\cite[p. 450]{ch}.

Moreover for Laplace equation the ``initial data'' are given on a closed
domain (a topological transformation of circles), for the wave equation,
they are given on a curve that can be considered as a topological
transformation of an equilateral hyperbola, i.e., of a curve with tangent
lines of a given kind \cite{ch}.

Now we observe that for the ``initial data'' for wave equation given on a curve, we can
consider, from a mathematical point of view, both the points at the left
or at the right of the curve. This possibility generate the 
difference \cite{ch} between the Laplace and wave equations with
respect to the initial data problem.
In particular, as we see in the following of this paper, for one arbitrary function  given on an
arm of equilateral hyperbola, we do not have an unique determination
of a function satisfying the wave equation and assuming on the line the given values, as it 
happens for Laplace equation, but 
from these values we can determine the sum of the values of the function in two points that,
extending to hyperbolic plane the Euclidean symmetry about the circle, can
be defined as {\sf symmetric} with respect to hyperbola (Sect. \ref{APOLL}).

Taking into account the analogies and these differences, we now
``translate'' to wave equation, studied in the hyperbolic geometry, the
classical results obtained for Laplace equation, studied in the Euclidean
geometry, in the internal points of a circle \cite{ch}.
\section{Initial Data on an Arm of Equilateral Hyperbola}\label{4.2}
\begin{figure}
\begin{center}
\mbox{\includegraphics[width=15.0cm]{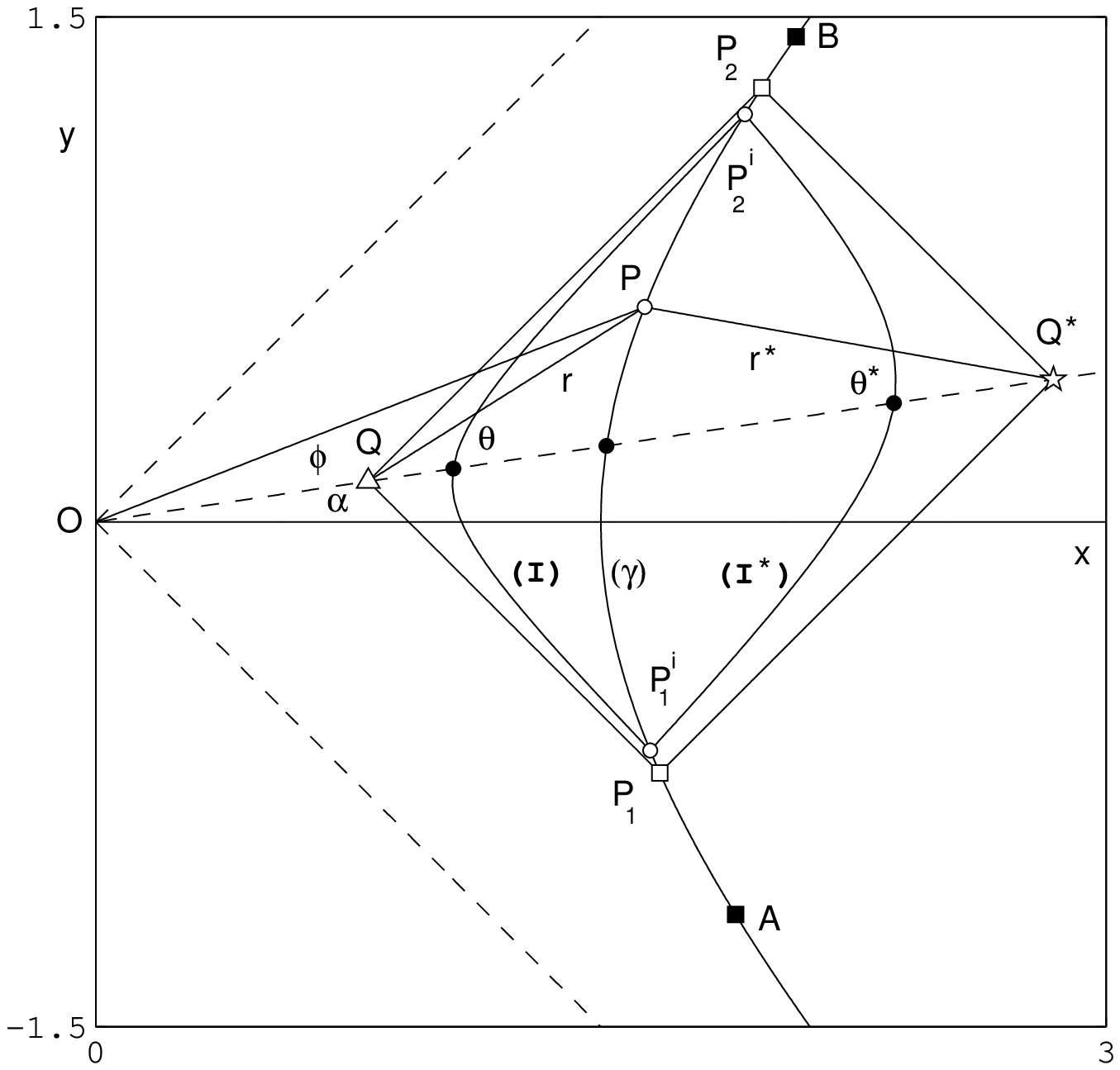}}
\caption
{{\bf The integration domains for wave equation in hyperbolic geometry.}\protect\\
In this figure geometric elements are represented for application of integral formulas to calculate
a function satisfying the wave equation, in the symmetric points $Q\,(\x,\,\y\,)$ and
$Q^*\,(\x^*,\,\y^*\,)$, with initial data given by an arbitrary function defined on 
the arc $AB$ of an equilateral hyperbola $\g$. \protect \\
$P_1$ and $P_2$ are the extreme points of 
the {\it domain of dependence} ( Sect. \ref{4.2}). In particular the following elements are reported \protect \\
$\bullet$ The equilateral hyperbolas $\g,\; \mathcal{I}$, $\mathcal{I}^*$, with semi-diameters $p,\,\r,\,\r^*$ and	their intersection points $P^i_1$ and $P^i_2$. 
\protect \\
$\bullet$ The hyperbolic angular variables $\f,\,\th,\,\th^*$,
measured with respect to the straight line $OQ^*$, set at a hyperbolic angle $\a$ with respect to x axis. \protect \\
$\bullet$ The hyperbolic distances $r$ and $r^*$ from a point $P$ of the hyperbola $\g$,
to points $Q$ and $Q^*$, respectively. \protect \\
Since the hyperbolas, in hyperbolic geometry, represent the locus of points at the same
distance from a given point, they correspond to circles of Euclidean geometry, used for
the same problems about Laplace equation studied in Euclidean plane. \protect \\ 
In particular, the hyperbolas $\mathcal{I}$ and $\mathcal{I}^*$, for which, in
the final step of the procedure, we do the limit $\r,\,\r^*\ra 0$,
correspond to the infinitesimal circles around the singularities. 
In this limit they become the parallel to axes bisectors from the points $Q$ and $Q^*$, respectively.
\label{fig3}}
\end{center}
\end{figure}

Let be given, as initial data, the values of one arbitrary function $u\,(\t)$
on the arc, between the points $A$ and $B$, of the right arm of an equilateral hyperbola
$\g$ with center in the axes origin $O$ and semi-diameter $p$ 
(Fig. \ref{fig3}). \\
We look for a function $u\,(\,x,\,y)$ satisfying the wave equation and assuming on $\g$
the given values. \\
By calling $Q\equiv (\x,\,\y)$ the point in which we look for $u\,(\,\x,\,\y)$, the parallels to axes
bisectors from $Q$ cross $\g$ in the points $P_1\,,P_2$. These points have to be internal
to hyperbola arc $AB$.	\\
In this way the
{\it domain of dependence}, determined by the point $Q$ and the hyperbola $\g$,
 is defined \cite[p. 438]{ch}. 

We draw from $P_1$ and $P_2$ the parallel to axes bisectors in the opposite direction with
respect to $Q$ and call $Q^*$ their intersection point. \\
We define $Q$ and $Q^*$ as {\it symmetric with respect to hyperbola $\g$},
by extending to hyperbolas in the hyperbolic plane, the Euclidean
symmetry with respect to a circle. Actually these points, as it is shown in \cite{Cl},
have equivalent properties:
\ben
\item they are on the same straight line through the center of hyperbola $\g$;
\item the product of their hyperbolic distances
from the center is equal to $p^2$ (squared semi-diameter).
\een
Let us set 
$Q^*\equiv (\x^*,\,\y^*)$ and call
\be
\overline{OQ}\equiv\sqrt{\x^2-\y^2}=q\,,  \label{ro0}
\ee
it results
\be
\x^*=\frac{p^2}{q^2}\,\x,\quad\y^*=\frac{p^2}{q^2}\,\y\,. \label{coostar}
\ee

The other two hyperbolas $\mathcal{I}$ and $\mathcal{I}^*$ with centers in $Q$ and $Q^*$,
represented in Fig. \ref{fig3}, are taken so that they
intersect each other on the hyperbola $\g$ in the points $P^i_1$ and $P^i_2$.
We call their semi-diameters $\r$ and $\r^*$, respectively. \\
For the other elements of  Fig. \ref{fig3}, we call
 $\a$  the hyperbolic angle between $x$ axis and the straight line $\overline{OQ}$ and
$\f,\,\th$, $\th^*$ the angles describing the hyperbolas, measured with respect to the
straight line $\overline{OQ}$. \\
The equations of the three hyperbolas, in  hyperbolic polar form, are given by
\ba
\g\,\,&\ra&\,\,x=p\,\cosh (\f+\a)\,;\qquad   y=p\,\sinh (\f+\a)\,  \label{ipegamma}    \\
\mathcal{I}\,\,&\ra&\,\,x=\x+\r\,\cosh (\th+\a)\,;\qquad  y=\y+\r\,\sinh (\th+\a)\,,
\label{ipetta}	\\
\mathcal{I}^*\,\,&\ra&\,\,x=\x^*-\r^*\,\cosh (\th^*+\a)\,;\qquad
y=\y^*-\r^*\,\sinh (\th^*+\a)\,.  \label{ipettastar}
\ea
Moreover we consider a point $P\equiv(x,\,y)$ 
and the distances $r=\overline{QP}$ and $r^*=\overline {Q^*P}$	given by
\be
r=\sqrt{(x-\x)^2-(y-\y)^2}\,;\qquad r^*=\sqrt{(x-\x^*)^2-(y-\y^*)^2}\,,  \label{rerstar}
\ee
where $\x^*$ and $\y^*$ are given by Eqs. (\ref{coostar}).

In Fig. \ref{fig4} we report the elements of Fig. \ref{fig3}  rotated 
by a hyperbolic angle $-\a$.  \\In this way a symmetric representation with respect
to $x$ axis is obtained. This representation allows a better insight in the geometric properties
and an easier way for definitions and calculations.

After this rotation, we have
\bit
\item the hyperbola $\g$ remains in the same position;
\item  all the hyperbolic distances and hyperbolic angles between corresponding lines are preserved;
\item the  hyperbolas $\mathcal{I}$ and $\mathcal{I}^*$ become symmetric with respect to x axis.
\eit
Thus, by calling $\f^i,\,\th^i,$ and $\th^{i*}$ the absolute values of the limit hyperbolic
angles for a point moving along the aforesaid hyperbolas from $P^i_1$ to $P^i_2$, the
ranges for the hyperbolic angles in Eqs. (\ref{ipegamma}-\ref{ipettastar}), are
\be
{\rm for}\,\, \g: \,\,\,-\f^i<\f<+\f^i\,,\,\,\,
{\rm for}\,\, \mathcal{I}: \,\,\,-\th^i<\th<+\th^i\,,\,\,\,
{\rm for}\,\, \mathcal{I}^*: \,\,\,+\th^{i*}>\th^*>-\th^{i*}\,.   \label{ranges}
\ee

\subsection{Extension of Apollonius Circle Theorem to Hyperbolic Plane}
\label{APOLL}
\begin{figure}
\begin{center}
\mbox{\includegraphics[width=15.0cm]{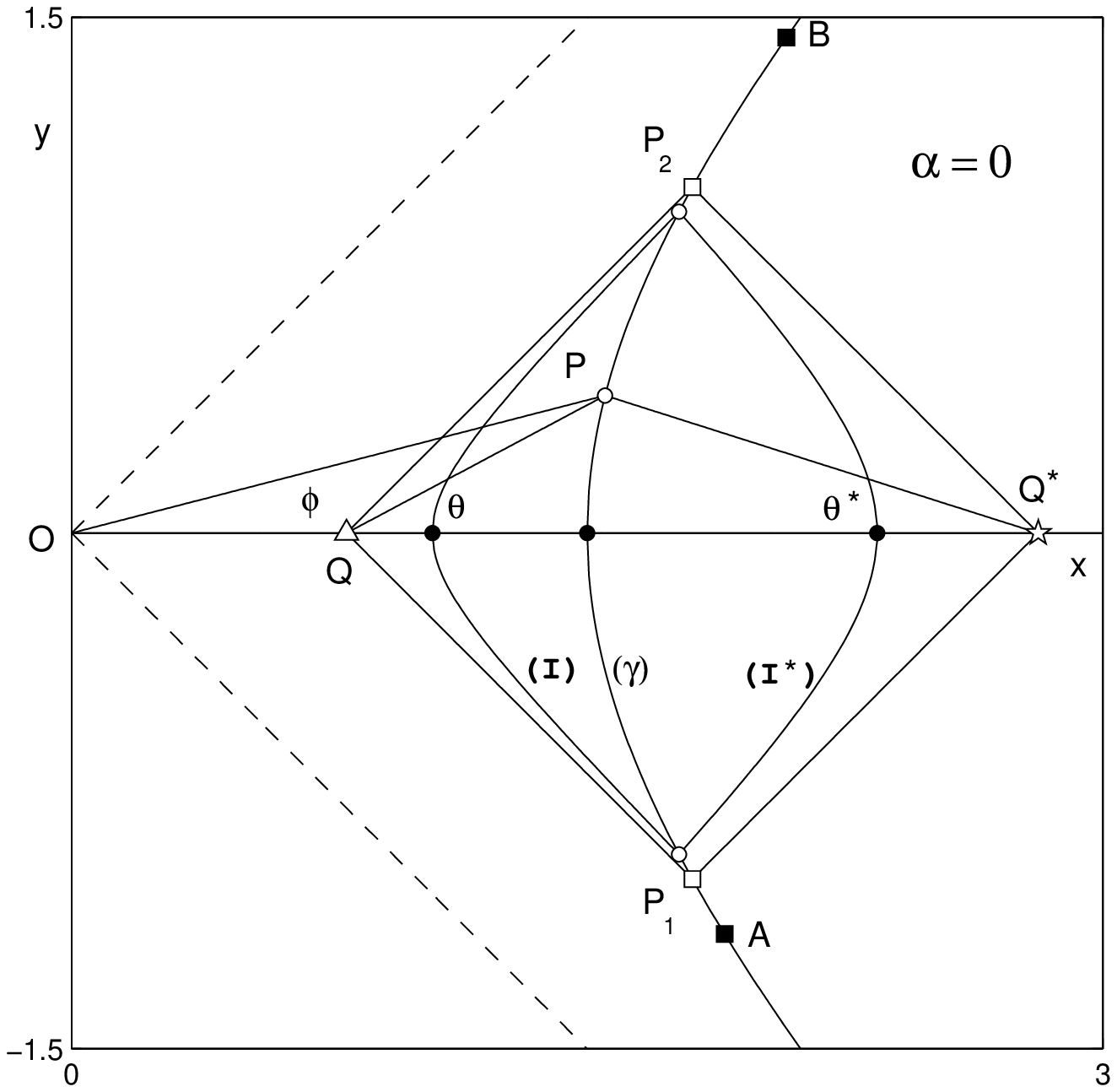}}
\caption
{{\bf Domain of dependence in symmetric position.}\protect\\
The coordinates of the points of Fig. \ref{fig3} are
transformed by means of a hyperbolic rotation of an angle $-\,\a$ in order to set 
$Q$ on the $x$ axis. In this way a symmetric representation with respect
to $x$ axis is obtained. This representation allows a better insight in the geometric properties
and an easier way for definitions and calculations.
 \protect\\ 
After this rotation, we have
 \protect\\ $\bullet$
the hyperbola $\g$ remains in the same position;
 \protect\\ $\bullet$
 all the hyperbolic distances and hyperbolic angles between corresponding lines are preserved;
 \protect\\ $\bullet$ the  hyperbola $\mathcal{I}$ and $\mathcal{I}^*$ become symmetric with respect to x axis.
 \protect\\ $\bullet$ 
the extreme points A and B of the arc on which the {\it initial data}	$u\,(\t)$, are given,
do not change.
\label{fig4}}
\end{center}
\end{figure}

We show that a similar theorem to the Apollonius theorem in Euclidean plane about the
circle, holds for equilateral hyperbolas in hyperbolic plane. \\
Referring to letters and symbols used in Fig. \ref{fig4}, we have \\
{\bf Theorem} - {\it Given a straight line from the center of a
hyperbola with semi-diameter $p$ and two points $Q$ and $Q^*$ on this
straight line, so that
\be \overline{OQ}\cdot \overline{OQ^*}=p^2 \Ra
\frac{\overline{OQ}}{p}=\frac{p}{\overline{OQ^*}},\label{Apo1} \ee}
{\it for all the points $P$ on the hyperbola,
we have} \be \frac{\overline{Q\,P}}{\overline{Q^*\,P}}=
\frac{\overline{OQ}}{p}.\label{Apo} \ee
This theorem can be demonstrated in a ``Euclidean way" thanks to the analytical formalization
of Hyperbolic geometry \cite{noi2}. Here we use the analytical approach that has allowed the recalled
formalization.

{\it Proof} - By using the definitions of the hyperbolic distances $r$ and $r^*$
given by Eq. (\ref{rerstar}) and the definition of $q$ given by Eq. (\ref{ro0}),
Eq. (\ref{Apo}) becomes
\be
\frac{r}{r^*}\equiv\frac{\sqrt{(x-\x)^2-(y-\y)^2}}{\sqrt{(x-\x^*)^2-(y-\y^*)^2}}=\frac{q}{p}\,.
\label{Aporrs}
\ee
By squaring 
this equation, it results
\be
\left(\frac{r}{r^*}\right)^2\equiv\frac{(x^2-y^2)+(\x^2-\y^2)-2(\x\,x-\y\,y)}
{(x^2-y^2)+(\x^{*2}-\y{*^2})-2(\x^*\,x-\y^*\,y)}=\frac{q^2}{p^2}\,.
\label{Apo2r}
\ee
By means of Eqs. (\ref{ro0}) and (\ref{coostar}), we see that Eq. (\ref{Apo2r}) is verified if
$P\equiv(x,\,y)$ belongs to $\g$, that is $x^2-y^2=p^2$.
Actually, in this case, we have
\be
\left(\frac{r}{r^*}\right)^2\equiv\frac{p^2+q^2-2(\x\,x-\y\,y)}
{p^2+\displaystyle\frac{p^4}{q^2}-2\displaystyle\frac{p^2}{q^2}(\x\,x-\y\,y)}=\frac{q^2}{p^2}\,.
\qquad\qquad\qquad \Box \label{Apo2pq}
\ee

For convenience we set
\be
\frac{q}{p}=A_p
\label{costap}
\ee
and Eq. (\ref{Aporrs}) becomes
\be
\left(\frac{r}{r^*}\right)_{P\,on\,\g}=A_p\,.	 \label{aplike}
\ee
From this theorem, by applying Eq. (\ref{aplike}) in the particular cases $P\equiv P^i_1$
or $P\equiv P^i_2$, it follows
\be
\frac{\r}{\r^*}\equiv\frac{\overline{QP^i_1}}{\overline{P^i_1 Q^*}}\equiv\frac{\overline{QP^i_2}}
{\overline{P^i_2 Q^*}}
\equiv\left(\frac{r}{r^*}\right)_{P\,on\,\g}=A_p\,.	 \label{aplikero}
\ee


\subsection{Application of Integral Formulas} \label{cicles}
Let us apply Eq. (\ref{ah5}) to the following domains represented in Fig. \ref{fig3}

\ben
\item between the hyperbolas $\mathcal{I}$ and $\g$;
\item between the hyperbolas  $\g$ and $\mathcal{I}^*$.
\een
Let $u\,(\,x,\,y\,)$ be a function that satisfies the wave equation in these domains. \\
In the application of Eq. (\ref{ah5}) to the domain $1$ we set
\be
v(x,\,y)=\ln r \,,  \label{Uln}
\ee
in the application to the domain $2$ we set
\be
v(x,\,y)=\ln r^* \,,  \label{Ulnstar}
\ee
where $r,\,r^*$ are given by Eq. (\ref{rerstar}). \\
It can be checked at once that these functions $v(x\,,y)$
satisfy the wave equation.   \\
From Eq. (\ref{ah5}) we have:  \\
for domain $1$
\be
\int^{P^i_2}_{P^i_1\,(\g)}\left(\ln r\,\frac{\partial u}{\partial n}-
u\,\frac{\partial \ln\, r}{\partial n}\right)\;d\t +
\int^{P^i_1}_{P^i_2\,(\mathcal{I})}\left(\ln r\,\frac{\partial u}{\partial n}-u\,\frac{\partial
\ln \,r}{\partial n}\right)\;d\t=0\,  \label{BGr1}
\ee
 for domain $2$
\be
\int^{P^i_1}_{P^i_2\,(\g)}\left(\ln r^*\,\frac{\partial u}{\partial n}-
u\,\frac{\partial \ln\, r^*}{\partial n}\right)\;d\t +
\int^{P^i_2}_{P^i_1\,(\mathcal{I}^*)}\left(\ln r^*\,\frac{\partial u}{\partial n}-u\,\frac{\partial
\ln \,r^*}{\partial n}\right)\;d\t=0\,.  \label{BGr11}
\ee
Let us add Eqs. (\ref{BGr1}) and (\ref{BGr11}) and group together the integrals on $\g$ and
the integrals on $\mathcal{I}$ and $\mathcal{I}^*$.  \\
As far as the integrals on $\g$ are concerned, for
Eq. (\ref{aplike}), it results $\ln\,r^*= \ln\,r-\ln\, A_p\,$.
Therefore we have
\ba
\int^{P^i_2}_{P^i_1\,(\g)}\ln r\,\frac{\partial u}{\partial n}\,d\,\t
-\int^{P^i_2}_{P^i_1\,(\g)}u\,\frac{\partial \ln\, r}{\partial n}\;d\t	
+\int^{P^i_1}_{P^i_2\,(\g)}\ln r\,\frac{\partial u}{\partial n}\,d\,\t -\nonumber \\
- \int^{P^i_1}_{P^i_2\,(\g)}\ln {A_p}\,\frac{\partial u}{\partial n}\,d\,\t
- \int^{P^i_1}_{P^i_2\,(\g)}u\,\frac{\partial \ln\, r^*}{\partial n}\;d\t\,.  \label{14inter}
\ea
The first and third integrals are equal, but have opposite integration direction
then their sum is zero. So, by collecting the second and the fifth integrals, Eq. (\ref{14inter})
becomes
\be
\int^{P^i_2}_{P^i_1\,(\g)}\ln A_p\,\frac{\partial u}{\partial n}\,d\,\t
-\int^{P^i_2}_{P^i_1\,(\g)}u\,\left[\frac{\partial \ln\,(r/{r^*})}{\partial n}\right]\;d\t\,.\label{14}
\ee
 \\
Now let us consider the integrals on $\mathcal{I}$ and $\mathcal{I}^*$.
They have to be calculated for $r=\r$ and $r^*=\r^*$, respectively.
From the relation (\ref{aplikero}), we have $\ln\,\r^*=\ln\,\r-\ln\,A_p\,$, therefore we have
\ba
\int^{P^i_1}_{P^i_2\,(\mathcal{I})}\ln \r\frac{\partial u}{\partial n}\,d\,\t
-\int^{P^i_1}_{P^i_2\,(\mathcal{I})}u\,\left(\frac{\partial \ln\,r}{\partial n}\right)
_{(r=\r)}\;d\t	 +\int^{P^i_2}_{P^i_1\,(\mathcal{I}^*)}\ln \r\,
\frac{\partial u}{\partial n}\,d\,\t +\nonumber   \\
+\int^{P^i_1}_{P^i_2\,(\mathcal{I}^*)}\ln A_p\,\frac{\partial u}{\partial n}\,d\,\t
-\int^{P^i_2}_{P^i_1\,(\mathcal{I}^*)}u\,\left(\frac{\partial \ln \,r^*}{\partial n}\right)
_{(r^*=\r^*)}\;d\t\,.\label{14a}
\ea
The sum of the first and third terms is an integral on a closed cycle and,
because $\ln\r$ is constant and $u$ satisfies the wave equation, it
follows, from Eq. (\ref{ah22}), that it is zero. 
For the same reason the integral on the closed cycle given by the sum between
the first term of Eq. (\ref{14}) and
the fourth one of Eq. (\ref{14a}) is equal to zero. \\
 After these reductions, the contribution of the derivative ${\partial u}/{\partial n}$  disappears
 and from the sum of Eqs. (\ref{BGr1}) and (\ref{BGr11})  remains
\be
\int^{P^i_2}_{P^i_1\,(\mathcal{I})}u\,\left(\frac{\partial \ln\,r}{\partial n}\right)
_{(r=\r)}\;d\t
+\int^{P^i_1}_{P^i_2\,(\mathcal{I}^*)}u\,\left(\frac{\partial \ln \,r^*}{\partial n}\right)
_{(r^*=\r^*)}\;d\t
=\int^{P^i_2}_{P^i_1\,(\g)}u\,\left[\frac{\partial \ln\,(r/{r^*})}{\partial n}\right]\;d\t\,.
\label{14resto}
\ee
\subsection{Introduction and Properties of $\mathcal{C}$ Function} \label{POli}
Let us now introduce a function $\mathcal{C}$ given by
\be
\mathcal{C}=\ln \left[\frac{p}{q}\left(\frac{r}{r^*}\right)\right]. \label{CF}
\ee
By using this function, the right hand side of Eq. (\ref{14resto}) may be written
\be
\int^{P^i_2}_{P^i_1\,(\g)}u\,\left[\frac{\partial \ln\,(r/{r^*})}{\partial n}\right]\;d\t\equiv
\int^{P^i_2}_{P^i_1\,(\g)}u\,\frac{\partial \mathcal{C}}{\partial n}\;d\t\;.  \label{usoC}
\ee
The $\mathcal{C}$ function has the following  properties	
\bit
\item  thanks to relation (\ref{aplike}), it  is zero on $\g$,
\item it satisfy the wave equation. \eit
Therefore it has the same  properties of Green function introduced for Laplace
equation \cite[Chap. 4]{ch}.
Here we see that, if the ''initial data" are given
on an arm of equilateral hyperbola, the $\mathcal{C}$ function, as the Green function,
allows us to obtain a Poisson-like integral.

\section{Poisson-like Integral}  \label{Poili}
The Poisson integral formula originates for solving in a circle
(and in a sphere), the ``initial data'' problem for two (and three)
dimensional Laplace equation and states a different way for tackle
the problem for elliptic partial differential equations (PDE), with
respect to Cauchy problem, about initial data (Sect. \ref{IDPr}).
It is known as ``Dirichelet problem'' \cite[p. 345]{sh}.
For two-dimensional Laplace equation it 
may be obtained in many ways \cite{ch} - \cite{sh}.
Here, studying the problem in hyperbolic plane, we obtain an analogous integral formula for
``initial data''  given on the right arm of equilateral hyperbola $\g$. We call it
{\sf Poisson-like integral, for the wave equation.}

\subsection{Poisson-like Kernel} \label{kernel}
With reference to Eq. (\ref{14resto})
let us consider the right hand side, in which the values of $u$ are given. \\
The first step is to calculate the kernel ${\partial\,\mathcal{C}}/{\partial\,n}$ of the integral.
From Eq. (\ref{CF}) it results
\be
\frac{\partial\,\mathcal{C}}{\partial\,n}=\frac{1}{r} \frac{\partial\,r}{\partial\,n}-\;
\frac{1}{r^*}\frac{\partial\,r^*}{\partial\,n}\,.   \label{pois01}
\ee
Let us carry out the derivatives  by considering $r$ and $r^*$, given by Eq. (\ref{rerstar}), as functions of $x,\,y$
\be
\frac{\partial\,r}{\partial\,n}=\frac{\partial\,r}{\partial\,x}\frac{\partial\,x}{\partial\,n}+
\frac{\partial\,r}{\partial\,y}\frac{\partial\,y}{\partial\,n}\,;\qquad
\frac{\partial\,r^*}{\partial\,n}=\frac{\partial\,r^*}{\partial\,x}\frac{\partial\,x}{\partial\,n}+
\frac{\partial\,r^*}{\partial\,y}\frac{\partial\,y}{\partial\,n}\,;	\label{idenpar}
\ee
Moreover, because the point $P\equiv(x,\,y)$ is on $\g$, it results   $$x^2-y^2=p^2\,,$$
and, from Eqs. (\ref{aplike}) and (\ref{chshsi}),
 the terms of the right hand side of Eq. (\ref{pois01}) become
\be
\frac{1}{r}\frac{\partial\,r}{\partial\,n}=\frac{1}{r}\left[(x-\x)\frac{x}{p}-
(y-\y)\frac{y}{p}\right]\equiv
\frac{1}{p\,r^2}\,[p^2-(\x\,x-\y\,y)]
\label{drC}
\ee
\be
\frac{1}{r^*}\frac{\partial\,r^*}{\partial\,n}=\frac{1}{r^*}\left[(x-\frac{p^2}{q^2}\,\x)\frac{x}{p}-
(y-\frac{p^2}{q^2}\,\y)\frac{y}{p}\right]\equiv
\frac{q^2}{p\,r^2}\,\left[1-\frac{1}{q^2}\,(\x\,x-\y\,y)\right]\,.
\label{drstarC}
\ee 
By substituting these expressions in relation (\ref{pois01}), after reduction, we obtain 
\be
\left(\frac{\partial\,\mathcal{C}}{\partial\,n}\right)_{on\,\g}
= \frac{p^2-q^2}{p\,r^2}\,. \label{po7}
\ee
Now we see that, by means of this expression, the integral in the right hand side of
Eq. (\ref{14resto}) gives a {\sf Poisson-like kernel}.

{\it  Proof} - By referring to Fig. \ref{fig3} 
let us calculate the hyperbolic distance $r=\overline{QP}$.
By applying the hyperbolic Carnot's theorem \cite{noi2} to the triangle
$\stackrel{\triangle}{QOP}$, where $\overline{OP}=p$, $\overline{OQ}=q$ and $\f$
is the hyperbolic angle $QOP$ 
\be
\overline{QP}^2\equiv r^2=p^2+q^{2}-2\,p\,q\cosh \f\,,
\label{rcarnot}
\ee
Moreover, on hyperbola $\g$ we have 
\be
d\t=p\,d\f\,.  \label{detaug}
\ee
Thus by means of Eqs. (\ref{po7}), (\ref{rcarnot})
and (\ref{detaug}), the right hand side of
Eq. (\ref{14resto}) becomes
\be
\int^{+\f^i}_{-\f^i\,(\g)}u\,(\,\f+\a)\,
\frac{p^2-q^2}{p^2+q^{2}-2\,p\,q\cosh \f}\;d\f \,.
\qquad\Box
\label{poissfinit}
\ee
Referring to Fig. \ref{fig3}, we note that as points $P_1^i$ and $P_2^i$ go toward the
extreme points $P_1$ and $P_2$, 
the sides $\overline{QP_1^i}$ and $\overline{QP_2^i}$ become parallel to axes bisectors 
and $r\ra 0$. Therefore, in this limit, the integral diverges.
\subsection{Limit of Poisson-like Integral on $\g$} \label{u1u2}
By referring to Fig. \ref{fig3}, we can note that as $P \ra P_1$ and $P_2$, more than $r\ra 0$,
we have $\th\ra\infty$. This angle is linked with the integration variable $\f$. \\
Here we see that the limits of the integral (\ref{poissfinit}) is of the  same order
as $\th^i\ra \infty$. This fact allows us to obtain for $r\ra 0$,  a finite result. \\

{\it Proof} - 
Let us divide the left and right sides of Eq. (\ref{14resto}) by $\th^i$ and let 
us begin by calculating the limit of the right-hand side
\be
\lim_{\,\th^{\,i}\,\ra\infty }\left[\frac{1}{2\,\th^{\,i}}\;
\int^{+\f^i}_{-\f^i\,(\g)}u\,(\,\f+\a)\,
\frac{p^2-q^2}{p^2+q^{2}-2\,p\,q\cosh \f}\;d\f\right]\,.
\label{POli2}	
\ee
This limit is an indeterminate form $\infty/\infty$.  \\
In order to calculate this limit, let us express the angle $\th^{\,i}$ as function of $\f^i$,
that is the limit value of the integral. In particular for the point $P_2^i \equiv
(p\,\cosh\f^i\,, p\,\sinh\f^i)$ of Fig. \ref{fig4}, we have
\be
\tanh\th^{\,i}=\frac{p\,\sinh\f_i}{p\,\cosh\f^i-q}\qquad\ra\qquad
\th^{\,i}=\tanh^{-1}\left[\frac{p\,\sinh\f_i}{p\,\cosh\f^i-q}\right]
\label{teidifii}
\ee
and, substituting in Eq. (\ref{POli2}), we obtain
\be
\lim_{\,\f^{\,i}\,\ra\f_2 }\left\{
\frac{\displaystyle\int^{+\f^i}_{-\f^i\,(\g)}u\,(\,\f+\a)\,\frac{p^2-q^2}{p^2+q^{2}-2\,p\,q\cosh \f}\;d\f}
{2\,\tanh^{-1}{\displaystyle\left[\frac{p\,\sinh\f^i}{p\,\cosh\f^i-q}\right]}}\;
\right\}\,.
\label{POli22}
\ee
 Let us apply the rule of L'Hospital by substituting to numerator and denominator their
derivatives with respect to $\f^i$. In this way  the integral in the numerator is eliminated
and we have
\be
\lim_{\,\f^{\,i}\,\ra\f_2 }\left\{\frac{
u\,(\,\f^i+\a\,){\displaystyle\frac{p^2-q^2}{p^2+q^{2}-2\,p\,q\cosh \f^i}}
+u\,(\,-\f^i+\a)\,{\displaystyle\frac{
p^2-q^2 }{p^2+q^{2}-2\,p\,q\cosh (-\f^i)}}  }
{  2\,{\displaystyle\frac{
p^2-p q \cosh\f^i} {p^2+q^{2}-2\,p\,q\cosh \f^i} }  }
\right\}\,.
\label{POli22H}
\ee
By calculating $\cosh\,\f_2$,
from the coordinates of extreme $P_2$ in Fig. \ref{fig4}, it results
$\cosh\f_2=(p^2+q^2)/(2\,p\,q)$ and we have
\be
\lim_{\,\f^{\,i}\,\ra\f_2 }\left[u\,(\,\f^i+\a)+u(\,-\f^i+\a)\right]=u\,(\,\f_2+\a)+u\,(\,-\f_2+\a)\,.
\equiv u\,(\,P_1)+u\,(\,P_2) \qquad  \Box
\label{POli22HF}
\ee

Actually we can say that the limit of the ratio between the Poisson kernel and $2\,\th^{\,i}$
acts as a {\it hyperbolic delta function}. In fact
the final result in Eq. (\ref{POli22HF}) is that the integral disappears and just the sum of
the values of the integrand calculated in the points $P_1$ and $P_2$ remains.
These points are the ones connected, by the parallel to axes bisectors, with the points
in which we are looking for the field.
\subsection{Limits of the Integrals on $\mathcal{I}$ and $\mathcal{I}^*$} \label{uustar}
Now let us consider the two terms of the left hand side of Eq. (\ref{14resto})
and calculate the same limits of the right hand side. \\
Let us express the two integrals
as functions of the hyperbolic angular variables $\th$ and $\th^{*}$, respectively,
and consider the local axis $n$
in the points of ${\mathcal{I}}$. It is directed like $r$ and,
referring to the right arm in Fig. \ref{fig1}, we observe that, because in Eq. (\ref{14resto})
the integration direction is upward, $n$ is oriented in the $r$ increasing direction
and ${\partial r}/{\partial n}=1$. We have
\be
\frac{\partial \ln r}{\partial n}\equiv
\left(\frac{\partial \ln r}{\partial r}\,\frac{\partial r}{\partial n}\right)_{r=\r}
=\frac{1}{\r}\,.      \label{delog}
\ee
Moreover, in the points of ${\mathcal{I}}$ it results $d\t=\r\,d\th$.  \\
From these positions and by recalling Eq. (\ref{ranges}) that gives the
ranges of the hyperbolic angles, the first term
of the left hand side of Eq. (\ref{14resto}) becomes
\be
\int^{P^i_2}_{P^i_1\,(\mathcal{I})}u\,\left(\frac{\partial \ln\,r}{\partial n}\right)
_{(r=\r)}\;d\t=
\int^{+\th^{\,i}}_{-\th^{\,i}\,(\mathcal{I})}u\,(\r,\,\th+\a)d\th\,.	\label{uzeroth}
\ee
Analogous considerations allow us to transform the integral on $\mathcal{I}^*$. \\
Let us consider the local axis $n$ in the points of $\mathcal{I}^*$. It is directed like $r^*$ and,
referring to the left arm in Fig. \ref{fig1}, we observe that, because the integration direction is
 downward, $n$ is oriented in the $r^*$ increasing direction,
then ${\partial r^*}/{\partial n}=1$ and we have
\be
\frac{\partial \ln r^*}{\partial n}\equiv
\left(\frac{\partial \ln r^*}{\partial r^*}\,\frac{\partial r^*}{\partial n}\right)_{r^*=\r^*}
=\frac{1}{\r^*}\,.	\label{delogst}
\ee
Moreover, in the points of $\mathcal{I}^*$ it results $d\t=\r^*\,d\th^*$.  \\
From these positions and by recalling Eq. (\ref{ranges}), the second term
of the left hand side of Eq. (\ref{14resto}) becomes
\be
\int^{P^i_1}_{P^i_2\,(\mathcal{I}^*)}u\,\left(\frac{\partial \ln\,r^*}{\partial n}\right)
_{(r^*=\r^*)}\;d\t=
\int^{+\th^{\,i*}}_{-\th^{\,i*}\,(\mathcal{I}^*)}u\,(\,\r^*,\th^*+\a)\,d\th^*\,.	  \label{uzerothst}
\ee
The values of $\th^{\,i}$ and $\th^{\,i*}$ are calculated from the intersection points
of $\mathcal{I}$ and $\mathcal{I}^*$ with $\g$, for $\a=0$ (Fig. \ref{fig4}).
By taking into account Eqs. ({\ref{ro0}) and (\ref{Apo1}) and setting
\be
q^*\equiv\overline{OQ^*}=\frac{p^2}{q}\,, \label{qstar}
\ee
it results
\ba
\cosh\th^{\,i}=\frac{p^2-q^2-\r^2}{2\,q\,\r}\,,\qquad \sinh\th^{\,i}=\frac{\sqrt{(\,p^2-q^2)^2
+\r^2(\,\r^2-2\,p^2-2\,q^2\,)}}{2\,q\,\r}\,, \label{lim}	   \\
\cosh\th^{\,i*}=\frac{-p^2+q^{*2}+\r^{*2}}{2\,q^*\,\r^*}\,,
\qquad \sinh\th^{\,i*}=\frac{\sqrt{(\,p^2-q^{*2})^2
+\r^{*2}\,(\,\r^{*2}-2\,p^2-2\,q^{*2})}}{2\,q^*\,\r^*}\,.
\label{limiti}
\ea
Let us demonstrate, referring to the particular case of Fig. \ref{fig4} ($\a=0$), that
\ba
\lim_{\,\th^{\,i}\,\ra\infty }\left[\frac{1}{2\,\th^{\,i}}\;
\int^{+\th^{\,i}}_{-\th^{\,i}\,(\mathcal{I})}u\,(\r,\,\th)\,d\th\right] = u\,(Q) \label{uzero} \\
\lim_{\,\th^{\,i}\,\ra\infty }\left[\frac{1}{2\,\th^{\,i}}\;
\int^{+\th^{\,i*}}_{-\th^{\,i*}\,(\mathcal{I}^*)}u\,(\r^*,\,\th^*)\,d\th^*\right] = u\,(Q^*)
\label{ustzero}
\ea

{\it Proof.}
Let us start from Eq. (\ref{uzero}) and write the function $u\,(\,\r,\,\th\,)$ on the points of hyperbola ${\mathcal{I}}$
by means of the Taylor's formula of order $1$
\ba
u\,(\,\r,\,\th)&\equiv& u\,(\,q+\r\,\cosh \th\,;\,\r\,\sinh \th)= u\,(\,Q\,) \nonumber \\
&+&\left(\frac {\partial u}{\partial x}\right)_Q \r\,\cosh \th+
\left(\frac {\partial u}{\partial y}\right)_Q\r\,\sinh \th
+\frac{1}{2!}\left[\left(\frac {\partial^2 u}{\partial x^2}\right)_{\,Q'}\r^2\cosh^2 \th\right. \nonumber \\
&+&\left. 2\left(\frac {\partial^2 u}{\partial x \partial y}\right)_{Q'} \r^2\cosh \th\sinh \th
+\left(\frac {\partial^2 u}{\partial y^2}\right)_{Q'}\r^2\sinh^2 \th\right]\,,	  \label{Taylor}
\ea
where $Q\equiv(q,\,0)$ and $Q'$ is an appropriate point on the segment between $Q$ and the
 point determined by $\r,\,\th$. \\
Let us substitute Eq. (\ref{Taylor}) in  the left hand side of	Eq. (\ref{uzero}).
Some terms do not give contribution, because they are anti-symmetric functions integrated
in a symmetric range. It results
\ba
\lim_{\,\th^{\,i}\,\ra\infty }\left\{\frac{1}{2\,\th^{\,i}}\,[2\,\th^{\,i}\,u(Q)]+\frac{1}{2\,\th^{\,i}}\,\left[
2\left(\frac {\partial u}{\partial x}\right)_Q \r\,\sinh\th^{\,i}\right.\right.  \nonumber \\
\left.\left.+\frac{1}{2}\left(\frac {\partial^2 u}{\partial x^2}\right)_{Q'}\r^2(\sinh \th^{\,i}\cosh \th^{\,i}+\th^{\,i})
+\frac{1}{2}\left(\frac {\partial^2 u}{\partial y^2}\right)_{Q'}\r^2(\sinh \th^{\,i}\cosh \th^{\,i}-\th^{\,i})
\right]\right\}\,. \label{limthi}
\ea
Let us express the hyperbolic functions of $\th^i$ as function of $\r$ by using Eqs. (\ref{lim}). After this substitution
the limit must be changed as follows $\displaystyle
\lim_{\,\th^{\,i}\,\ra\,\infty}\ra
\lim_{\,\r\,\ra\,0}\,. $    \\
As far as the terms in square brackets are concerned in this limit, we have 
\bit
\item the hyperbolic functions are proportional to $1/\r$, so the 
products between powers of $\r$ and the same power of hyperbolic functions give finite values;
\item the terms $\r^2\,\th^{\,i}$ go to zero.
\eit
Then only $u\,(Q)$ remains and Eq. (\ref{uzero}) is obtained.

The demonstration of Eq. (\ref{ustzero}) is obtained by means of analogous considerations, taking into account that
$\displaystyle\lim_{\,\r\,\ra\,0}\,\r^*=0.\,$ But in the final
step we divide for the divergent range $2\,\th^{\,i}$ that is different from $2\,\th^{\,i*}$
at numerator. Now we  demonstrate that, in the limit $\r\,\ra\,0$, the two divergent ranges are equal. \\
Actually from Eqs. (\ref{costap}), (\ref{aplikero}) and (\ref{qstar}),
let us	express $\cosh\th^{\,i*}$ as function of  the same parameters of $\cosh\th^{\,i}$.
By applying the rule of L'Hospital, we obtain
\be
\lim_{\,\th^{\,i}\,\ra\,\infty}\left(\frac{2\,\th^{\,i*}}{2\,\th^{\,i}}\right)\equiv\lim_{\,\r\,\ra\,0}\left\{\frac{2\,\cosh^{-1}{\displaystyle\left[
\frac{p^2-q^2+\r^2}{2\,p\,\r}\right]}}{2\,\cosh^{-1}{\displaystyle\left[
\frac{p^2-q^2-\r^2}{2\,q\,\r}\right]}}
\right\}=1\,. \qquad\Box  \label{limrho}
\ee

Let us extend the demonstration to the general case, in which $\a\ne 0$. So let us demonstrate,
referring to Fig. \ref{fig3}, that
\ba
\lim_{\,\th^{\,i}\,\ra\infty }\left[\frac{1}{2\,\th^{\,i}}\;
\int^{+\th^{\,i}}_{-\th^{\,i}\,(\mathcal{I})}u\,(\r,\,\th+\a)\,d\th\right] = u\,(Q) \label{uzeroalfa} \\
\lim_{\,\th^{\,i}\,\ra\infty }\left[\frac{1}{2\,\th^{\,i}}\;
\int^{+\th^{\,i*}}_{-\th^{\,i*}\,(\mathcal{I}^*)}u\,(\r^*,\,\th^*+\a)\,d\th^*\right] = u\,(Q^*)
\label{ustzeroalfa}
\ea

{\it Proof.}
Let us start from Eq. (\ref{uzeroalfa}), the Taylor formula of Eq. (\ref{Taylor})
for hyperbola $\mathcal{I}$, must be generalized as follows
\ba
u\,(\r,\,\th+\a)&\equiv& u\,(\x+\r\,\cosh \,(\th+\a)\,;\,\y+\r\,\sinh \,(\th+\a))= u\,(Q) \nonumber \\
&+&\left(\frac {\partial u}{\partial x}\right)_Q \r\,\cosh \,(\th+\a)+
\left(\frac {\partial u}{\partial y}\right)_Q\r\,\sinh\, (\th+\a)
+\frac{1}{2!}\left[\left(\frac {\partial^2 u}{\partial x^2}\right)_{Q'}\r^2\cosh^2 (\th+\a)\right. \nonumber \\
&+&\left. 2\left(\frac {\partial^2 u}{\partial x \partial y}\right)_{Q'} \r^2\cosh\, (\th+\a)\sinh \,(\th+\a)
+\left(\frac {\partial^2 u}{\partial y^2}\right)_{Q'}\r^2\sinh^2 (\th+\a)\,\right]\,.	 \label{Taylorg}
\ea
where $Q\equiv(\x,\,\y)$ and $Q'$ is an appropriate point on the segment between $Q$ and the point determined
by $\r,\,\th+\a$. \\
So, in the calculation of definite integrals, $\sinh\th^i$ and $\cosh\th^i$ are substituted
by linear combinations of the same functions and all the previous considerations about the limit for
 $\r\, \ra\,0$ hold. \\
The same generalization can be done for Eq. (\ref{ustzeroalfa}) and hyperbola $\mathcal{I^*}$.
$\qquad\qquad\qquad\Box$  \\

By collecting the results of Eqs. (\ref{uzeroalfa}), (\ref{ustzeroalfa}) and (\ref{POli22HF}),
it results
\be
u\,(Q)+u\,(Q^*)=u\,(P_1)+u\,(P_2)\,.  \label{final}
\ee

As a final remark, we give a mathematical meaning to the integrals in
Eqs. (\ref{uzero}) and (\ref{ustzero}) divided by the diverging angle $2\th^i$. \\
Actually, from Eq. (\ref{uzero}) we see that the factor $1/ (2\,\th_i)$ outside the integral,
is the same of the integration limits, then the left hand side represent the mean value of the
function $u$ in the integration range. Taking into account Eq. (\ref{limrho}), the same meaning
can be given to  Eq. (\ref{ustzero}).  \\
This result is in agreement with the equivalent expressions in the studies of functions of a
 complex variable \cite{sh} and  for  ``the initial value problem"  for Laplace equation 
 \cite{ch}.\\ Actually for these problem, for 
calculating the value of a function in a point $P$ of a domain, given its values on the frontier, it  is
 taken a circle around $P$ with radius $r\ra 0$. This circle  gives a factor $1/(2\, \pi)$
 before the integrals of the right hand side,  calculated in the  range $0\leftrightarrow 2\, \p$.

\section{Non-omogeneous Wave Equation} \label{nonomo}
Here we see that the classical approach, by means of integral formulas of Sec. \ref{IH},
allows us to extend the obtained results to non-omogeneous wave equation.  \\
Let $v$ satisfy the omogeneous wave equation and $u$ the non-omogeneous one, that is
\be
\D_2\,u=f\,(x,\,y)\,.  \label{nonomequ}
\ee
Eq. (\ref{ah5}) becomes
\be
\oint_\G\left(v\,\frac{\partial u}{\partial n}-u\,\frac{\partial v}{\partial n}
\right)\;d\t=\int_D\int\,v\,f(x,\,y)\,dx\,dy  \label{nonah5}
\ee
and, by setting in Eq. (\ref{ah4}) $v=1$, Eq. (\ref{ah22})
becomes
\be
\oint_\G \frac{\partial u}{\partial n}\;d\t=\int_D\int\,f(x,\,y)\,dx\,dy\,.
\label{nonah22}
\ee
\subsection{Application of Integral Formulas}  \label{noncicles}
Let us do the same steps that are done in Sec. \ref{cicles} for the omogeneous wave equation.
Referring to Fig. \ref{fig3}, let us apply Eq. (\ref{nonah5}) to the following domains
\ben
\item domain $D$ between the hyperbolas $\mathcal{I}$ and $\g$;
\item domain $D^*$ between the hyperbolas  $\g$ and $\mathcal{I}^*$.
\een
Let $u\,(\,x,\,y\,)$ be a function that satisfies the non-omogeneous wave equation
(\ref{nonomequ}) in these domains. \\
In the application of Eq. (\ref{nonah5}) to the domain $D$ we set
\be
v(x,\,y)=\ln r \,,  \label{nonUln}
\ee
in the application to the domain $D^*$ we set
\be
v(x,\,y)=v^*(x,\,y)=\ln r^* \,,  \label{nonUlnstar}
\ee
where $r,\,r^*$ are given by Eq. (\ref{rerstar}). \\
It can be checked at once that these functions $v(x\,,y)$
satisfy the wave equation.   \\
From Eq. (\ref{nonah5}) we have:  \\
for domain $D$
\ba
\int^{P^i_2}_{P^i_1\,(\g)}\left(\ln r\,\frac{\partial u}{\partial n}-
u\,\frac{\partial \ln\, r}{\partial n}\right)\;d\t +
\int^{P^i_1}_{P^i_2\,(\mathcal{I})}\left(\ln r\,\frac{\partial u}{\partial n}-u\,\frac{\partial
\ln \,r}{\partial n}\right)\;d\t=  \nonumber \\
\int_D\int(\,\ln r)\,f(x,\,y)\,dx\,dy\,,  \label{nonBGr1}
\ea
for domain $D^*$
\ba
\int^{P^i_2}_{P^i_1\,(\g)}\left(\ln r^*\,\frac{\partial u}{\partial n}-
u\,\frac{\partial \ln\, r^*}{\partial n}\right)\;d\t +
\int^{P^i_1}_{P^i_2\,(\mathcal{I^*})}\left(\ln r^*\,\frac{\partial u}{\partial n}-u\,\frac{\partial
\ln \,r^*}{\partial n}\right)\;d\t=  \nonumber \\
\int_{D^*}\int(\,\ln\,r^*)\,f(x,\,y)\,dx\,dy\,.  \label{nonBGr11}
\ea

Let us add the left and the right-hand sides of  Eqs. (\ref{nonBGr1}) and (\ref{nonBGr11})
and group together the
integrals on $\g$ and the integrals on $\mathcal{I}$ and $\mathcal{I}^*$.  \\
From this sum the following total equation results, which includes the integrations on the two domains
$D$ and $D^*$
\ba
\int^{P^i_2}_{P^i_1\,(\g)}\ln A_p\,\frac{\partial u}{\partial n}\,d\,\t
-\int^{P^i_2}_{P^i_1\,(\g)}u\,\left[\frac{\partial \ln\,(r/{r^*})}{\partial n}\right]\;d\t \nonumber \\
+\int^{P^i_1}_{P^i_2\,(\mathcal{I})}\ln \r\frac{\partial u}{\partial n}\,d\,\t
-\int^{P^i_1}_{P^i_2\,(\mathcal{I})}u\,\left(\frac{\partial \ln\,r}{\partial n}\right)
_{(r=\r)}\;d\t	 +\int^{P^i_2}_{P^i_1\,(\mathcal{I}^*)}\ln \r\,
\frac{\partial u}{\partial n}\,d\,\t \nonumber	 \\
+\int^{P^i_1}_{P^i_2\,(\mathcal{I}^*)}\ln A_p\,\frac{\partial u}{\partial n}\,d\,\t
-\int^{P^i_2}_{P^i_1\,(\mathcal{I}^*)}u\,\left(\frac{\partial \ln \,r^*}{\partial n}\right)
_{(r^*=\r^*)}\;d\t  \nonumber \\
=\int_D\int(\,\ln r)\,f(x,\,y)\,dx\,dy+\int_{D^*}\int(\,\ln\,r^*)\,f(x,\,y)\,dx\,dy\,.
\label{non14}
\ea
In the equation the left-hand side is the sum of Eqs. (\ref{14}) and (\ref{14a}), for the same
considerations done in Sec. \ref{cicles}. \\
By examining the sum of line integrals in the left-hand side, it results that the third and fifth
terms make up an integral on the frontier of domain $D+D^*$ and the first and sixth terms make up
an integral on the frontier of domain $D^*$. By applying Eq. (\ref{nonah22}) to these
closed cycles it results
\be
\int^{P^i_1}_{P^i_2\,(\mathcal{I})}\ln \r\frac{\partial u}{\partial n}\,d\,\t
+\int^{P^i_2}_{P^i_1\,(\mathcal{I}^*)}\ln \r\,\frac{\partial u}{\partial n}\,d\,\t
=\ln\r\int_{D+D^*}\int\,f(x,\,y)\,dx\,dy\,, \label{cicl1}
\ee
\be
\int^{P^i_2}_{P^i_1\,(\g)}\ln A_p\,\frac{\partial u}{\partial n}\,d\,\t
+\int^{P^i_1}_{P^i_2\,(\mathcal{I}^*)}\ln A_p\,\frac{\partial u}{\partial n}\,d\,\t
=-\ln A_p\int_{D^*}\int\,f(x,\,y)\,dx\,dy   \label{cicl2}
\ee
and, Eq. (\ref{non14}) becomes
\ba
\int^{P^i_2}_{P^i_1\,(\mathcal{I})}u\,\left(\frac{\partial \ln\,r}{\partial n}\right)
_{(r=\r)}\;d\t
+\int^{P^i_1}_{P^i_2\,(\mathcal{I}^*)}u\,\left(\frac{\partial \ln \,r^*}{\partial n}\right)
_{(r^*=\r^*)}\;d\t
-\int^{P^i_2}_{P^i_1\,(\g)}u\,\left[\frac{\partial \ln\,(r/{r^*})}{\partial n}\right]\;d\t \nonumber \\
=-\ln\r\int_{D+D^*}\int\,f(x,\,y)\,dx\,dy+\ln A_p\int_{D^*}\int\,f(x,\,y)\,dx\,dy  \nonumber \\
+\int_D\int(\,\ln r)\,f(x,\,y)\,dx\,dy+\int_{D^*}\int(\,\ln\,r^*)\,f(x,\,y)\,dx\,dy\,.
\label{reorgan}
\ea
The last equation is the extension of Eq. (\ref{14resto}) to the case of non-omogeneous wave equation.
\subsection{Limits of Area Integrals}  \label{areainteg}
In Secs. \ref{u1u2} and \ref{uustar} the terms of the left-hand side of Eq. (\ref{reorgan})
are divided by the integration range $2\,\th^{\,i}$ where, from Eq. (\ref{lim}),
\be
\th^{\,i}=\cosh^{-1}\left[\frac{p^2-q^2-\r^2}{2\,q\,\r}\right] \label{tetiaco}
\ee
and the limit of the ratio is calculated for $\r\ra 0$ (i.e., $2\,\th^{\,i}\ra \infty$).
In this way the results of Eqs. (\ref{uzeroalfa}), (\ref{ustzeroalfa}) and (\ref{POli22HF})
are obtained.	 \\
Here the same procedure is applied to the terms of the right-hand side of Eq. (\ref{reorgan}),
in order to extend the Eq. (\ref{final}) to the case of non-omogeneous wave equation.
When $\r\ra 0$ the hyperbolas $\mathcal{I}$ and $\mathcal{I}^*$ become the parallels to
axes bisectors from $Q$ and $Q^*$ (see Sec. \ref{GREEN-La} and Fig. \ref{fig3}) and the domains
$D$ and $D^*$ are contained by the arc $P_1 P_2$ of hyperbola $\g$ and the straight line
segments $QP_1,\,QP_2$ and $Q^*P_1,\,Q^*P_2$, respectively.

The following limit values result for the four area integral terms  of Eq. (\ref{reorgan})
\ben
\item
\be
\lim_{\,\r\,\ra\,0}\left\{\frac{\displaystyle -\ln\r\int_{D+D^*}\int\,f(x,\,y)\,dx\,dy}
{\displaystyle 2\cosh^{-1}\left[\frac{p^2-q^2-\r^2}{2\,q\,\r}\right]}\right\}
=\frac{1}{2}\int_{D+D^*}\int\,f(x,\,y)\,dx\,dy\,.
\label{primo}
\ee
{\it Proof} - The result of Eq. (\ref{primo}) is obtained by applying the rule of L'Hospital to
calculate
\be
\lim_{\,\r\,\ra\,0}\left\{\frac{\displaystyle -\ln\r}
{\displaystyle 2\cosh^{-1}\left[\frac{p^2-q^2-\r^2}{2\,q\,\r}\right]}\right\}=\frac{1}{2}\,.
\label{primopro}
\ee
Also it may be obtained by substituting the expression
\be
\cosh^{-1}\left[\frac{p^2-q^2-\r^2}{2\,q\,\r}\right]\equiv
\ln\left[\left(\frac{p^2-q^2-\r^2}{2\,q\,\r}\right)+
\sqrt{\displaystyle\left(\frac{p^2-q^2-\r^2}{2\,q\,\r}\right)^2-1}\right]
\label{argcosh}
\ee
and calculating the limit directly.$\qquad\qquad\qquad\qquad\qquad\qquad\qquad\qquad\qquad\qquad\Box$ \\
\item
\be
\lim_{\,\r\,\ra\,0}\left\{\frac{\displaystyle \ln A_p\int_{D^*}\int\,f(x,\,y)\,dx\,dy}
{\displaystyle 2\cosh^{-1}\left[\frac{p^2-q^2-\r^2}{2\,q\,\r}\right]}\right\}
=0\,.
\label{secondo}
\ee
{\it Proof} - Eq. (\ref{secondo}) is obtained by the direct limit calculation
\be
\lim_{\,\r\,\ra\,0}\left\{\frac{\displaystyle \ln A_p}
{\displaystyle 2\cosh^{-1}\left[\frac{p^2-q^2-\r^2}{2\,q\,\r}\right]}\right\}=0\,.
\qquad\qquad\qquad\qquad\qquad\qquad\Box
\label{secondopro}
\ee
\item
\be
\lim_{\,\r\,\ra\,0}\left\{\frac{\displaystyle \int_D\int(\,\ln r)\,f(x,\,y)\,dx\,dy}
{\displaystyle 2\cosh^{-1}\left[\frac{p^2-q^2-\r^2}{2\,q\,\r}\right]}\right\}
=0\,.
\label{terzo}
\ee
{\it Proof} - Let $M$ be the absolute maximum of $|f\,(x,\,y)|$
in the domain $D$ at limit $\r\ra 0$. Let us transform the coordinates $x,\,y$ into polar coordinates
\be
x=\x+r\,\cosh(\th+\a)\,;\qquad\qquad y=\y+r\,\sinh(\th+\a)\,.	\label{polari}
\ee
The area element is transformed
\be
dx\,dy=r\,dr\,d\th\,.	\label{elemarea}
\ee
It results
\be
\left|{\int_D\int(\,\ln r)\,f(x,\,y)\,dx\,dy}\right|< M
\left|\int^{(p-q)}_{\r}(\,\ln r)\,r\,dr\int^{\th^{\,r}}_{-\th^{\,r}}d\,\th\right|\,,
\label{maggzn}
\ee
where $\pm\,\th^{\,r}$ are the extreme values of $\th$, corresponding to the intersections points
of the hyperbola that has center in $Q$ and semi-diameter $r$, with hyperbola $\g$. The hyperbolic
angle$\th^{\,r}$ is given by Eq. (\ref{tetiaco}) or (\ref{argcosh}), where $\r$ is substituted by
$r$. Then, by using the logaritmic form (\ref{argcosh}),
\ba
&&\left|{\int_D\int(\,\ln r)\,f(x,\,y)\,dx\,dy}\right|< \nonumber \\
&&2M\left|\int^{(p-q)}_{\r}(\,\ln r)\,r\,\left\{\ln\left[\left(\frac{p^2-q^2-r^2}{2\,q\,r}\right)+
\sqrt{\displaystyle\left(\frac{p^2-q^2-r^2}{2\,q\,r}\right)^2-1}\right]\right\}d\,r\right|\,.
\label{maggrho}
\ea
The function in the integral in the right hand side has finite values for $\r\le r\le(p-q)$ and,
when $r=\r$ and $\r\ra 0$,
\ba
\lim_{\,r\,\ra\,0}\left\{(\,\ln r)\,r\,\ln\left[\left(\frac{p^2-q^2-r^2}{2\,q\,r}\right)+
\sqrt{\displaystyle\left(\frac{p^2-q^2-r^2}{2\,q\,r}\right)^2-1}\right]\right\} \nonumber \\
\quad\ra\quad
\lim_{\,r\,\ra\,0}\left\{(\,\ln r)\,r\,\ln\left[\frac{p^2-q^2}{\,q\,r}\right]\right\}
\quad\ra\quad \lim_{\,r\,\ra\,0}\left\{(\,\ln r)\,r\,(-\ln r)\right\}=0 \,.
\label{limitato}
\ea
The result of Eq. (\ref{limitato}) is obtained by applying two times the rule of L'Hospital.
Therefore the function in the area integral at numerator of (\ref{terzo}) is finite in all
the domain $D$, also in the limit $\r\ra 0$. Then the corresponding area integral is finite and,
divided by the hyperbolic angular range that diverges when $\r\ra 0$, gives the result of
Eq. (\ref{terzo}).
$\qquad\qquad\qquad\qquad\Box$
\item

\be
\lim_{\,\r\,\ra\,0}\left\{\frac{\displaystyle \int_{D^*}\int(\,\ln\,r^*)\,f(x,\,y)\,dx\,dy}
{\displaystyle 2\cosh^{-1}\left[\frac{p^2-q^2-\r^2}{2\,q\,\r}\right]}\right\}
=0\,.
\label{quarto}
\ee
{\it Proof} - A process of demonstration analogous to the one developed for Eq. (\ref{terzo})
can be used, recalling that $\displaystyle\lim_{\,\r\,\ra\,0}\,\r^*=0\,.$
$\qquad\qquad\qquad\qquad\qquad\qquad\qquad\qquad\qquad\qquad\qquad\qquad\Box$
\een
Finally, by collecting the results of Eqs. (\ref{uzeroalfa}), (\ref{ustzeroalfa}), (\ref{POli22HF}) and
(\ref{primo}), (\ref{secondo}), (\ref{terzo}), (\ref{quarto}), the following equation
is obtained
\be
u\,(Q)+u\,(Q^*)=u\,(P_1)+u\,(P_2)+\frac{1}{2}\int_{D+D^*}\int\,f(x,\,y)\,dx\,dy\,,
\label{nonfinal}
\ee
that is the extension of Eq. (\ref{final}) to the case of non-omogeneous wave equation.

\section{Conclusions}	 \label{conc}
By studying the wave equation in a plane with its own geometry, similar results to the ones of
Laplace equation studied in Euclidean plane are obtained.

These results can also be read in the following way: in \cite{noi2} the Euclidean theorems 
have been used as starting points for their extension to hyperbolic geometry by means of 
analytical demonstrations. In this paper, this method is extended to a problem,
related to wave equation that is usually studied in Euclidean geometry. \\
We know that the wave equation is the starting point for obtaining the Lorentz transformation 
\cite{booklet} from which a physical meaning to hyperbolic geometry is given. Therefore 
the application of hyperbolic geometry can be considered a natural way to study the wave equation
 in a Cartesian plane.

Taking into account the theoretical and practical
relevance of wave equation in Mathematics and Physics and the many
subjects related to the treated arguments, it can be expected that the
obtained results may be the starting point for improvements
in many directions.

\appendix
\section{Normal and Tangent Local Coordinates in the Hyperbolic Plane}	\label{ntau}
\begin{figure}
\begin{center}
\mbox{\includegraphics[width=15.0cm]{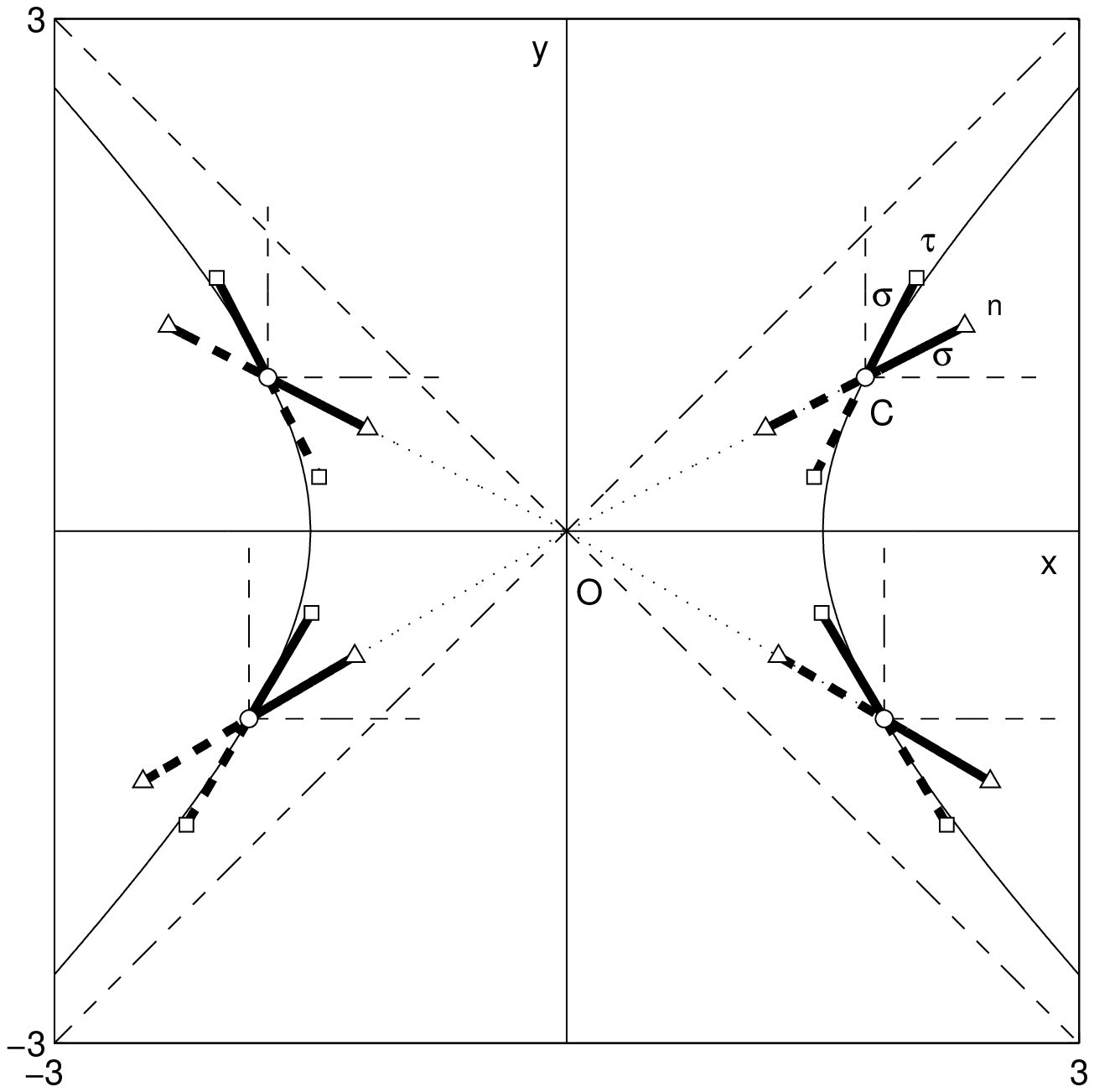}}
\caption
{{\bf Normal ($n$) and tangent ($\t$) unity vectors along a hyperbola in the hyperbolic plane.}
\protect \\ The topology and geometry of hyperbolic plane generate different definitions
from the ones of Euclidean geometry, now we recall \cite{noi2} and represent the orthogonal
lines. \protect \\
We call $\si$ the angle between the $\t$ axis and the parallel to $y$ axis of the local 
reference frame.\protect  \\
In this figure we show how normal ($n$) and tangent ($\t$) unity vector pairs
are defined, with respect to $\t$ orientation, in the points of both the arms of a hyperbola
and for $y>0$ or $y<0$. \protect \\
The solid lines represent $\t$ unity vector in the upward direction.
The $n$ unity vector is in the right sector of the local
reference frame. $\si$ is positive when
$\t$ unity vector is on the right hand side with respect to the positive local Cartesian $y$ axis
and is negative when it is on the left hand side. \protect  \\
The dashed lines represent $\t$ unity vector in the downward direction.
The $n$ unity vector is in the left sector of the local reference frame. $\si$ is positive when
$\t$ unity vector is on the left hand side with respect to  the negative local Cartesian y axis
and is negative when it is on the right hand side.  \label{fig1}}
\end{center}
\end{figure}

Here we construct the local reference frame used for calculating line integrals 
in a hyperbolic plane. \\
This reference frame has the origin in the points that move along the curve
and an axis tangent to the curve. It is equivalent to the local reference frame
employed in the line integrals in  Euclidean plane. Otherwise
the different topology and metric properties \cite{noi2} generate different
geometric characteristics.  \\
Calling $C(x_c,y_c)$ the generic point on the curve, a local Cartesian
reference frame is defined with origin in $C$. The coordinates on the local
axes, parallel to $x$ and $y$ axes, are $x-x_c$ and $y-y_c$. \\
In this reference, let us consider another reference, taking a	
$\t$ axis  tangent to the curve and oriented according with the integration direction.
It forms a hyperbolic angle $\si$ with respect to $y$ local axis.
The other axis, $n$,
forms the same hyperbolic angle $\si$ with respect to $x$
local axis. $n$ and $\t$ axes are orthogonal in the hyperbolic geometry, i.e., they
are symmetric with respect to the axes bisectors 
 of the local Cartesian reference frame. These axes are oriented so that the $n,\,\t$
 frame is congruent with the $x,\,y$ frame. \\
In Fig. \ref{fig1} an equilateral hyperbola is represented
and four positions of the point $C$ are considered. For each position the two possible
directions of $\t$ axis are reported. In this way the types of pairs of
unity vectors defining $n,\t$ reference axes are described. 

The transformation equations that link the coordinates $x,y$ to the coordinates $n,\t$ 
are
\ba
x-x_c&=&\pm\,\,[n\,\cosh\,\si+\t\,\sinh\,\si]	   \nonumber   \\
y-y_c&=&\pm\,\,[n\,\sinh\,\si+\t\,\cosh\,\si]	   \label{lorena}
\ea
with the inverse one
\ba
n&=&\pm\,\,[(x-x_c)\,\cosh\,\si-(y-y_c)\,\sinh\,\si]	      \nonumber   \\
\t&=&\pm\,\,[-(x-x_c)\,\sinh\,\si+(y-y_c)\,\cosh\,\si]\,.     \label{lorenb}
\ea
In these equations
the sign $+$ is used when $\t$ is oriented upward and the sign $-$ when $\t$ is
oriented downward.

Let us consider a function $u$ defined in the plane $x,\,y$.
By using Eqs. (\ref{lorena}) and (\ref{lorenb}) let us express, as a function of the
local variables $n,\,\t$, the differential form
\be
\frac{\partial u}{\partial x}\;d\,y+\frac{\partial u}{\partial y}\;dx  \label{term}
\ee
By considering a line integral of this form, 
it results $dn=0$, so that
\ba
dx&=&\frac{\partial x}{\partial n}\;dn+\frac{\partial x}{\partial \t}\;d\t\equiv
\frac{\partial x}{\partial \t}d\t  \nonumber  \\
dy&=&\frac{\partial y}{\partial n}\;dn+\frac{\partial y}{\partial \t}\;d\t\equiv
\frac{\partial y}{\partial \t}\;d\t\,.	\label{dexdey}
\ea
On the other hand, from Eq. (\ref{lorena}), the following relations between
the derivatives with respect to the orthogonal directions
of the tangent and the ``hyperbolic normal'' to the curve result
\be
\frac{\partial\,x}{\partial\,\t}= \pm\,\sinh\,\si=\frac{\partial \,y}{\partial \,n}\qquad
\frac{\partial\,x}{\partial \,n}= \pm\,\cosh\,\si=\frac{\partial \,y}{\partial \,\t}\,
\label{lorecr}
\ee
and the differential form of  Eq. (\ref{term}) becomes
\be
\frac{\partial u}{\partial x}\;d\,y+\frac{\partial u}{\partial y}\;dx
=\left(\frac{\partial u}{\partial x}\;\frac{\partial y}{\partial \t}
+\frac{\partial u}{\partial y}\;\frac{\partial x}{\partial \t}\right)\;d\t
\equiv \left(\frac{\partial u}{\partial x}\;\frac{\partial x}{\partial n}
+\frac{\partial u}{\partial y}\;\frac{\partial y}{\partial n}\right)\;d\t
=\frac{\partial u}{\partial n}\;d\t.  \label{equi}
\ee
\\

{\bf Partial Derivative with Respect to Normal Local Coordinates.}   \\ 

Referring to Fig. \ref{fig1}, partial derivatives of $x$ and $y$ with respect to $n$
are calculated with the conditions
\bit
\item the point of coordinates $x,\,y$ is on the equilateral hyperbola $\g$ with center in $O$
and semi-diameter $p$. Its equation can be represented by Eq. (\ref{ipegamma}), by setting
the line parameter  $\f+\a=\si$. 
\item $\t$ is oriented upward.
\eit
From Eqs. (\ref{lorena})  in which $+$ sign is used and Eqs. (\ref{ipegamma}), we have
\be
\frac{\partial x}{\partial n}=\cosh\,\si\,\equiv \frac{x}{p};\qquad \frac{\partial y}{\partial n}=\sinh\,\si\equiv \frac{y}{p}.
\label{chshsi}
\ee


\begin{thebibliography}{99}
\bibitem{noi2} {F. Catoni, R. Cannata, V. Catoni} and {P Zampetti, }
{\em  Hyperbolic Trigonometry in Two-dimensional Space-time Geometry},
{ Nuovo Cimento B}, {\bf 118 } (5), 475 (2003) (reprinted in
\cite{libro} and \cite{booklet})
\bibitem{libro} {F. Catoni, D. Boccaletti, R. Cannata, V. Catoni,
E. Nichelatti} and {P Zampetti,} {\em  The Mathematics of Minkowski
Space-Time}, {Birkh\"auser Verlag}, Basel (2008);
\bibitem{booklet} {F. Catoni, D. Boccaletti, R. Cannata, V. Catoni
} and {P Zampetti,} {\em  Geometry of Minkowski
Space-Time}, Springer-Verlag, Heidelberg (2011) 
\bibitem{Cl} {F. Catoni,  P Zampetti,} {\em Cauchy-like Integral
Formula for Functions of a Hyperbolic Variable}, Advances in Applied 
Clifford Algebra  {\bf 22}, 23 (2012)
\bibitem{ch} R. Courant and D. Hilbert, {\em Methods of Mathematical
 Physics}, Vol II, Interscience Publishers (1962).
 \bibitem{mf} P. M. Morse and H. Fesbach, {\em Methods of Theoretical Physics},
Mc Graw-Hill, New York (1953).
\bibitem{ST} {A. Sveshnikov, A. Tikhonov}, {\em The Theory of
Functions of a Complex Variable}, Mir, Moscou (1978).
\bibitem{sh}  Y.V. Sidorov, M.V. Fednyuk and M.I. Shabunin, {\em Lectures
on the Theory of Functions of a Complex Variable}, Mir, Moscou (1985).
\end{thebibliography}
\end{document}